# A novel strategy for achieving a low-field lightweight permanent MRI magnet system with good magnetic field homogeneity and low eddy current

Jingzhi Han (韩景智)[1#*], Jiangqian Guo（郭江黔）[2#], Peng Shen（沈鹏）[1,3,4,5], Xiao Tong（童骁）[1], Wenyun Yang（杨文云）[1], Ziheng Zhang (张子衡)[6,7], Jie Liu（刘杰）[2], Tianzhuo Yang（杨天卓）[1], Yikun Fang（方以坤）[3,4,5], Shunquan Liu（刘顺荃）[1], Jie Zhang（章杰）[2], Qing Xu（徐庆）[1], Jinbo Yang（杨金波）[1]

[1] Institute of Condensed Matter and Material Physics, School of Physics, Peking University, Beijing, 100871, China

[2] NingBo ChuanShanjia Electrical and Mechanical CO., LTD, Yuyao, 315699, China

[3] Division of Functional Materials, Central Iron & Steel Research Institute, Beijing, 100081, China

[4] Hefei Gangyan Rare Earth Permanent Magnetic Materials Research Institute Co., Ltd, Hefei, 230041, China.

[5]State Key Laboratory of Rare Earth Permanent Magnetic Materials; Hefei 230041, China.

[6] Beijing Ri-Bei Technology Co., Ltd., Beijing, 102299, China

[7] Beijing MagnVue Medix Co., Ltd. Beijing, 101111, China

[#]: Jingzhi Han and Jiangqian Guo have the same contribution.

*: corresponding author: hanjingzhi@pku.edu.cn

**Abstract:**

In low-field, lightweight, pole-pieceless permanent-magnet MRI systems built with sintered Nd-Fe-B or Sm-Co magnets, the rapid switching of gradient fields readily induces eddy currents in the sintered magnets, leading to image artifacts. To address this, we report for the first time a Sm-Fe-N permanent-magnet MRI system based on anisotropic Sm-Fe-N bonded magnets, whose high electrical resistivity reduces the eddy currents in the X, Y and Z directions to 0.93 ‰, 1.72 ‰ and 2.38 %, respectively, while a magnetic field inhomogeneity below 150 ppm is achieved at the boundary of a 220 mm diameter of spherical volume (DSV). Compared with sintered Nd-Fe-B and Sm-Co magnets, using Sm-Fe-N bonded magnets as the source of the static magnetic field not only suppresses eddy currents but also makes a closely tiled, densely packed magnetic-circuit layout feasible, providing a more uniform static magnetic field for the MRI system. Imaging results free of obvious geometric distortion and banding artifacts further indicate that the Sm-Fe-N magnet system delivers low eddy currents and high static magnetic field homogeneity.

## 1. Introduction

High-field MRI systems are widely preferred by patients because they are non-invasive, highly safe, and offer excellent spatial resolution.[1,2] However, their high purchase price and substantial maintenance cost leave developing countries and underdeveloped regions with very limited per-capita access. [3] The shortage of MRI systems often forces patients to wait a long time before they can be examined. In addition, high-field MRI systems are relatively bulky (weighing over 10000 kg) and difficult to move. To ensure timely diagnosis and treatment, the development of low-field, lightweight MRI systems has become particularly urgent.[3-12]

At present, the static magnetic field of low-field lightweight MRI systems is provided mainly by the magnetic circuit composed of sintered Nd-Fe-B magnets, sintered Sm-Co magnets, or a hybrid of Nd-Fe-B magnets and Sm-Co magnets.[13-19] These two sintered magnets offer a high maximum energy product, which favors lightweight low-field MRI design, but their electrical resistivity is very low. The rapid switching of gradient fields therefore readily induces eddy currents in the sintered Nd-Fe-B and Sm-Co magnets. To suppress these eddy currents, sintered Nd-Fe-B and Sm-Co magnets are commonly arranged into a segmented, relatively dispersed ring-shaped magnetic circuit. [19-20] Such a segmented layout, however, severely degrades static-field homogeneity, and because the eddy currents in sintered magnets cannot be completely eliminated, the parasitic field they generate perturbs the timing, waveform, and spatial linearity of the gradient field. This is particularly detrimental to gradient-timing-sensitive sequences, such as diffusion-weighted imaging (DWI) and echo-planar imaging (EPI).[3]

To address the eddy-current-induced degradation of image quality and static-field homogeneity in current low-field, lightweight, pole-pieceless permanent-magnet MRI systems based on sintered Nd-Fe-B and Sm-Co magnets, we propose for the first time the use of anisotropic Sm-Fe-N bonded magnets, which offer a much higher electrical resistivity, as the source of the static magnetic field. Based on the magnetic and electrical properties of the anisotropic Sm-Fe-N injection-molded magnets, the magnetic field of the Sm-Fe-N magnet system was simulated using COMSOL Multiphysics, leading to a magnetic-circuit design in which a 0.067 T static field is produced by a Sm-Fe-N magnet system with a 650 mm pole-disk diameter and a 300 mm pole-disk separation. Building on these calculations, the corresponding Sm-Fe-N magnet system was fabricated, and its static-field homogeneity, eddy-current behavior, and imaging performance were systematically evaluated.

## 2. Experimental Section

The Sm-Fe-N injection-molded magnets were supplied by Shenzhen High mag Technology Co., Ltd. The iron yokes, supports, and anti-eddy-current plates were supplied by Ningbo Chuanshanjia Electrical and Mechanical Co., Ltd. Considering both magnetic permeability and mechanical strength, A3 steel was selected for both the iron yoke and the supports; its saturation magnetic flux density was taken as 1.6 T in the simulations and calculations. COMSOL Multiphysics was used to simulate the static magnetic field of the magnet system. Based on the simulation results and the magnetic field orientation injection-molding process for the Sm-Fe-N magnets, the magnetic-circuit layout and the detailed geometry of the Sm-Fe-N magnets and the iron yoke were determined. On this basis, the Sm-Fe-N permanent-magnet MRI assembly was fabricated together with matched gradient coils, radio frequency coils, and receive coils. Gx and Gy gradient coils were unshielded, but Gz gradient coil was shielded in the single layer. The homogeneity and stability of the static field were investigated with a nuclear magnetic resonance (NMR) magnetometer and a Gauss meter. To evaluate the magnitude of eddy currents, a gradient magnetic field of 6 mT/m was applied for 1000 ms along the X, Y, and Z directions, respectively. Starting from 100 μs after the gradient field pulse ends, the linear term of the gradient field induced by the eddy current was measured along the direction of the applied gradient field. The magnitude of eddy current in the X, Y, and Z directions was calculated by dividing the measured the linear term of the gradient eddy current field by the corresponding applied gradient field (6 mT/m). In order to obtain the signal-to-noise Ratio (SNR), contrast-to-noise Ratio (CNR), spatial resolution, geometric distortion, image uniformity, and other parameters of the Sm-Fe-N magnet system, quantitative phantom validation on the magnet system were performed using American College of Radiology (ACR) MRI phantom (190 mm in diameter and 148 mm in height). The ACR phantom was positioned according to the ACR instructions. As recommended by the ACR guidelines, a localizer scan of the ACR phantom was first performed using a spin echo (SE) sequence. Subsequently, the ACR phantom was scanned using a T1-weighted SE sequence, a T2-weighted SE sequence, the routinely used T1-weighted sequence of the equipment, and the routinely used T2-weighted sequence of the equipment. Imaging experiments were performed on a spectrometer manufactured by Firstech.

## 3. Results and Discussion

The Sm-Fe-N bonded magnet used in this work was prepared by magnetic field orientation injection molding and the weight ratio of Sm-Fe-N magnetic powder in the magnet is 90 %. Figure 1 shows the X-ray diffraction pattern measured along the orientation direction together with the corresponding demagnetization curve. The (006) peak is the strongest in the XRD pattern, indicating pronounced anisotropy. Other peaks such as (321) are still visible in the oriented XRD pattern, indicating that the alignment of the Sm-Fe-N powder did not reach 100%. This is likely due to an insufficient orientation magnetic field, a relatively low molding temperature, or excessive friction between the magnetic powder and the mold during injection.

From the demagnetization curve the magnet has a remanence of 0.68 T, a coercivity of 7300 Oe, and a maximum energy product of 10.6 MGOe. The temperature coefficient of remanence of the magnet was also measured and found to be −0.07 %/°C, better than the −0.11 %/°C typical of Nd-Fe-B. [21-23] This indicates that, as the source of the static magnetic field in MRI magnet systems, Sm-Fe-N magnets exhibit a superior ability to resist temperature drift compared to Nd-Fe-B magnets.

The bulk resistance of injection-molded Sm-Fe-N magnets and sintered Nd-Fe-B magnets of identical geometry was compared. The Sm-Fe-N injection-molded magnet showed a bulk resistance of about 50 MΩ, whereas the sintered Nd-Fe-B magnet showed about 1 mΩ, i.e., the Sm-Fe-N magnet has a resistance roughly $10^{10}$ times higher than the sintered Nd-Fe-B magnet. This high resistance arises from the 10 wt% non-conductive PA12 binder used in the injection-molded magnet.

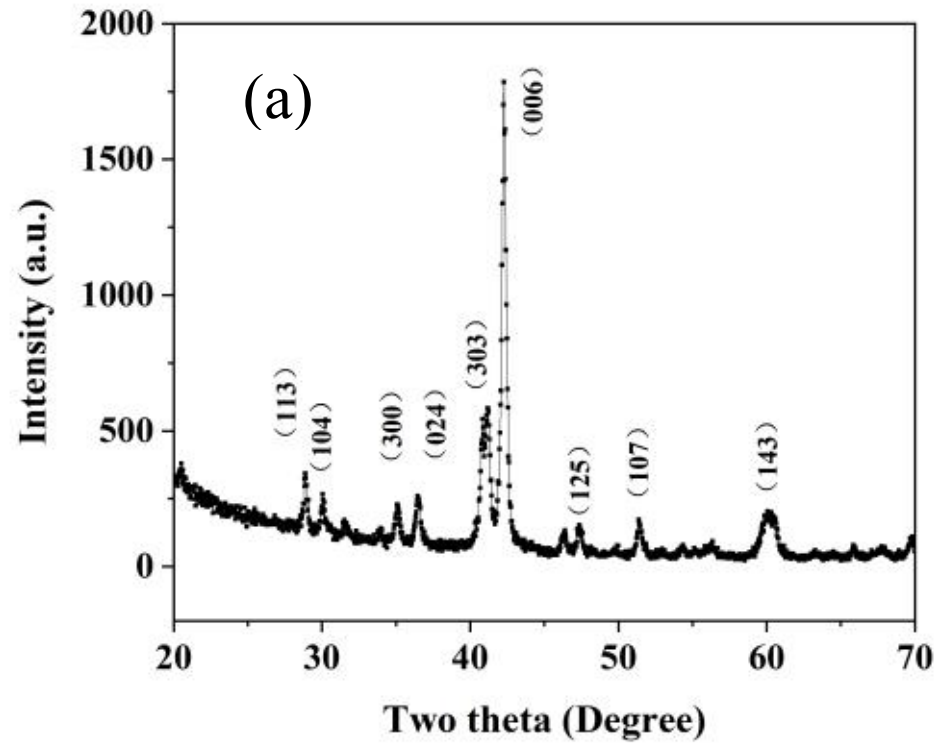

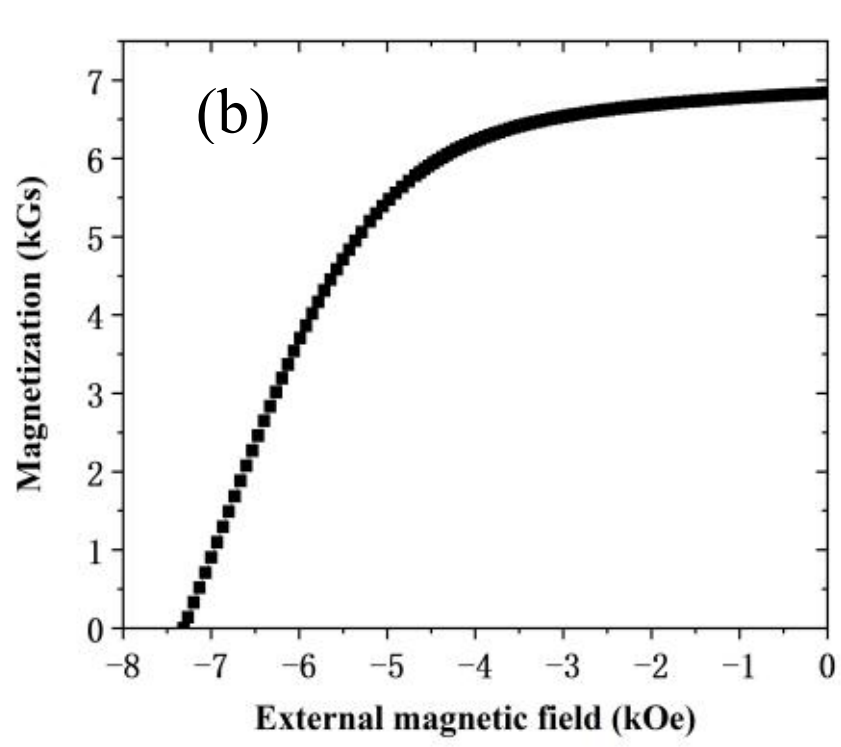

Figure 1. (a) Oriented X-ray diffraction pattern and (b) demagnetization curve of the Sm-Fe-N bonded magnet.

It is well established that, in MRI, the diameter of spherical volume (DSV) is closely related to imaging quality. To obtain a relatively large DSV with high internal magnetic field homogeneity while keeping the system weight reasonable, we designed and built a square-frame, low-field, lightweight Sm-Fe-N magnet system with a 650 mm pole-disk diameter and a 300 mm pole-disk separation. The target DSV was 220 mm at a field strength of 0.067 T.

During magnetic-circuit design, the very high bulk resistance of the Sm-Fe-N injection-molded magnet, which approaches that of an insulator, implies that the rapid switching of gradient fields produces only a very small eddy current inside the magnet. The Sm-Fe-N injection-molded magnets could therefore be arranged in a closely tiled, densely packed pattern on both the upper and lower pole disks to maximize static-field homogeneity, as shown in Figure 2.

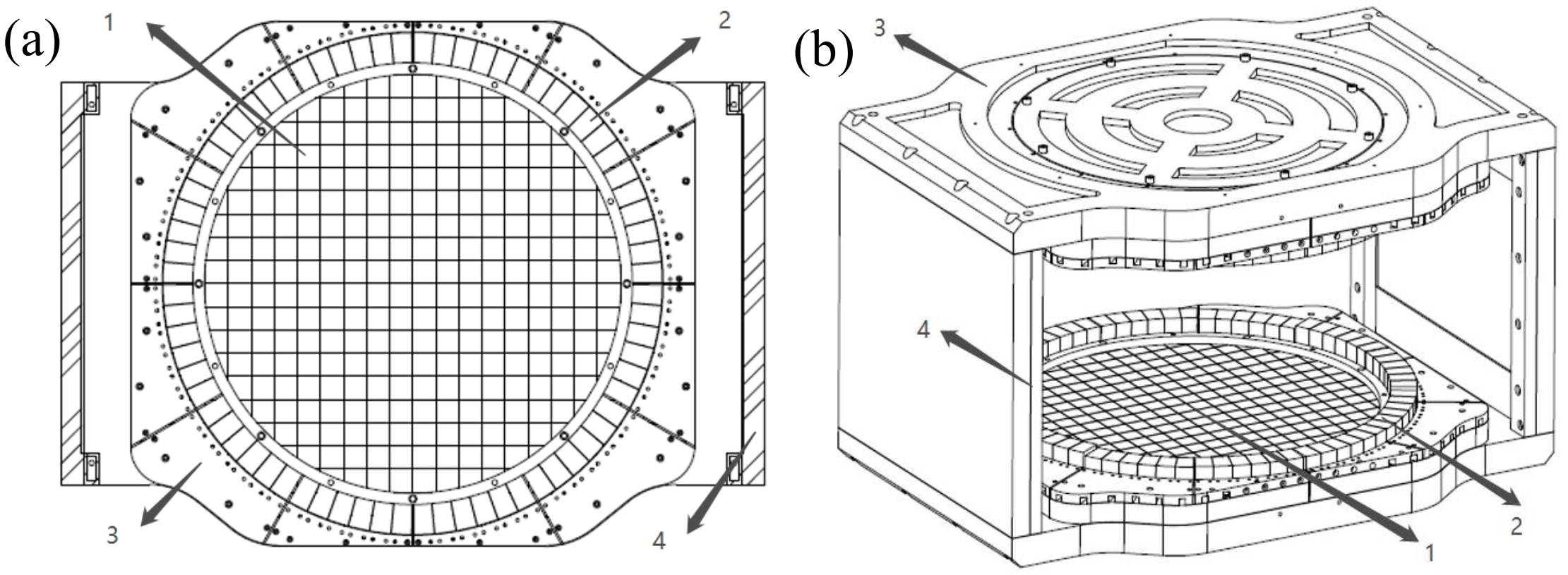

Figure 2. Schematic of the closely tiled, densely packed magnetic-circuit layout of (a) the Sm-Fe-N magnets on the pole disks and of (b) the overall magnet system structure. 1 is Magnetic-circuit layout in the central region of the pole disk; 2 is ring-shaped magnetic-circuit segment at the periphery; 3 is pole-disk support; 4 is iron yoke and supports.

A3 steel was used for the yoke and supports connecting the upper and lower pole disks (Figure 2). It was selected for two reasons: its relatively high magnetic permeability concentrates the magnetic flux and increases the magnetic field strength, while its mechanical strength ensures the structural stability of the magnet system. To prevent eddy currents from forming in the A3-steel yoke and supports during gradient operation, anti-eddy-current plates were installed between the upper yoke and the upper pole disk and between the lower yoke and the lower pole disk.

Building on this magnetic-circuit design, the static field of the Sm-Fe-N magnet system was simulated in COMSOL Multiphysics using the magnetic properties of the injection-molded magnets and the A3 steel together with the geometric dimensions of the magnet system. To produce a 0.067 T static magnetic field, the simulations indicated the following layout. Taking the geometric centers of the upper and lower pole disks as the origin, Sm-Fe-N magnets 30 mm thick fill a circular region 572.2 mm in diameter.

At the periphery of each pole disk, the Sm-Fe-N magnets form a ring with a 650 mm outer diameter and a 572.2 mm inner diameter. Within this ring, the magnets are 67.5 mm thick on the left and right sides next to the supports and 63 mm thick on the front and back sides away from the supports. The greater magnet thickness at the periphery relative to the central region, and at the support-side edges relative to the front- and back-side edges, was introduced to improve static-field homogeneity.

Following the magnetic-circuit layout obtained from the simulations, and combined with the magnetic field-oriented injection-molding process for the Sm-Fe-N bonded magnets, three types of injection-molded magnets meeting the design specifications were produced; their dimensions are detailed in Figure S1 of the Supplementary Material. The thickness of the injection-molded magnets along the orientation direction was kept within 11 mm to ensure a high degree of alignment of the Sm-Fe-N powder during magnetic field-oriented injection molding under a fixed orientation field. These three magnet types were assembled into the magnetic circuit as designed; the assembly procedure is described in the Supplementary Material. The completed Sm-Fe-N magnet system is shown in Figure 3. The Sm-Fe-N injection-molded magnets used in the system weigh approximately 143 kg, while the iron yokes, supports, anti-eddy-current plates and other accessories weigh approximately 400 kg, giving a total system mass of 543 kg. It is relatively light compared to traditional midfield and high-field permanent-magnet MRI magnet systems, which typically weigh several tons, so the MRI magnet system constructed with Sm-Fe-N injection magnets qualifies as a lightweight magnet system.

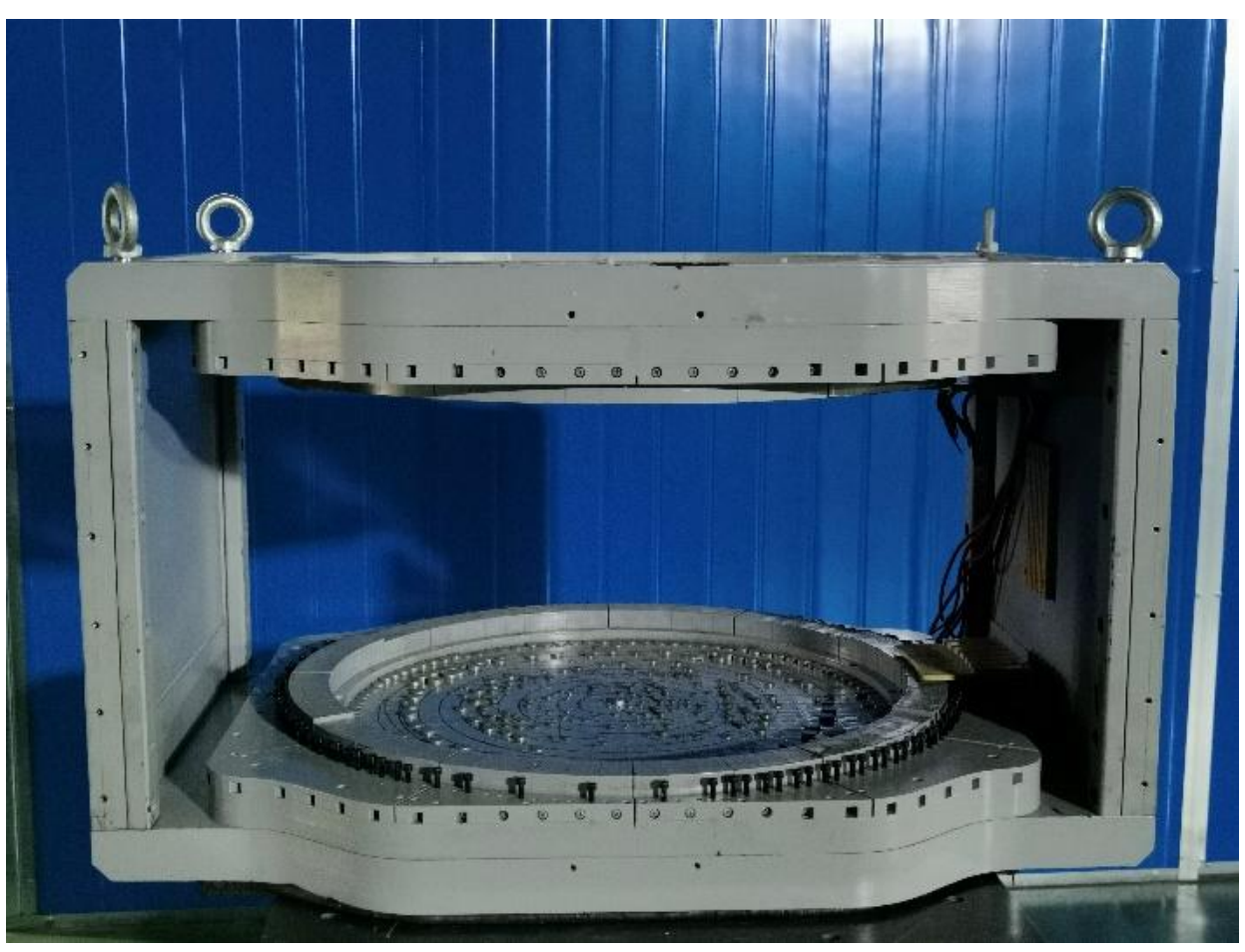

Figure 3. Photograph of the assembled Sm-Fe-N MRI magnet system. X direction: anterior-posterior direction; Y direction: left-right direction; Z direction: vertical direction.

The static magnetic field strength of the as-built Sm-Fe-N magnet system was measured first; within a spherical volume of 220 mm DSV the field is approximately 0.067 T (Figure 4). On the spherical surface, the magnetic field strength gradually increases from $Z_0$ to $Z_7$ and likewise from $Z_0$ to $-Z_7$. For an ideally homogeneous MRI magnet system, the field-variation amplitude from $Z_0$ to $+Z_7$ should equal that from $Z_0$ to $-Z_7$, but in the as-built system the two are slightly different, with the change from $Z_0$ to $-Z_7$ larger than that from $Z_0$ to $+Z_7$. This asymmetry mainly reflects the limited consistency of the dimensions of the Sm-Fe-N magnets and of their placement during assembly. Further passive shimming is therefore required to reach the high homogeneity sought in this Sm-Fe-N MRI magnet system.

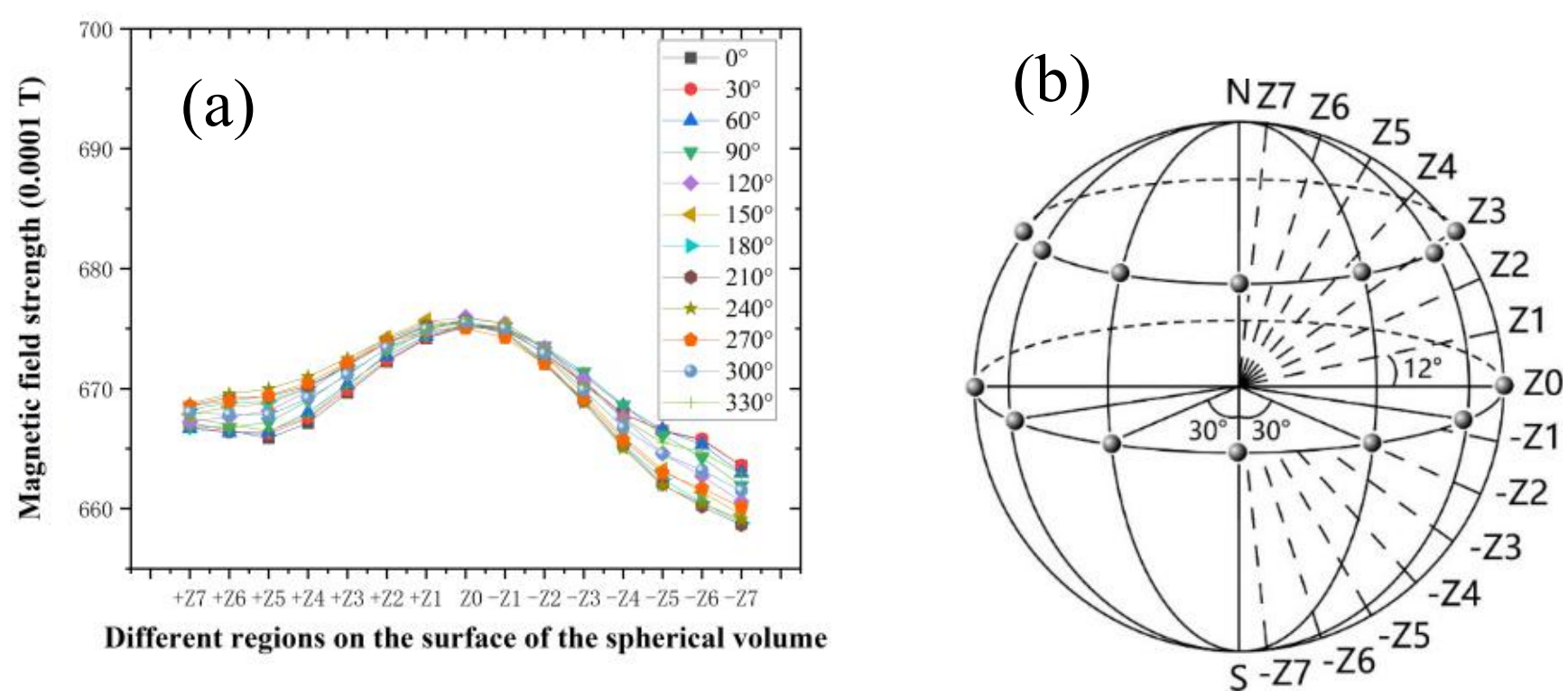

Figure 4. (a) Magnetic-field distribution on the surface of a sphere 220 mm in diameter centered on the geometric center of the magnet system. (b) Diagram of the sampling positions: at each $Z_i$ ($i = -7, -6, \ldots, 6, 7$), the field was measured every 30° along the circle defined by the intersection of the corresponding horizontal plane with the sphere. [19]

Passive shimming was then used to improve homogeneity. Small magnetic shims were placed at specific locations on the surface of the Sm-Fe-N magnets on both pole disks. By iteratively optimizing their positions and number, the field inhomogeneity within DSVs of 220 mm and 200 mm was reduced to 141.39 ppm and 92.62 ppm, respectively (Figure 5), which is superior to that of the hybrid magnet system composed of Sm-Co and Nd-Fe-B magnets.[19]

(a)

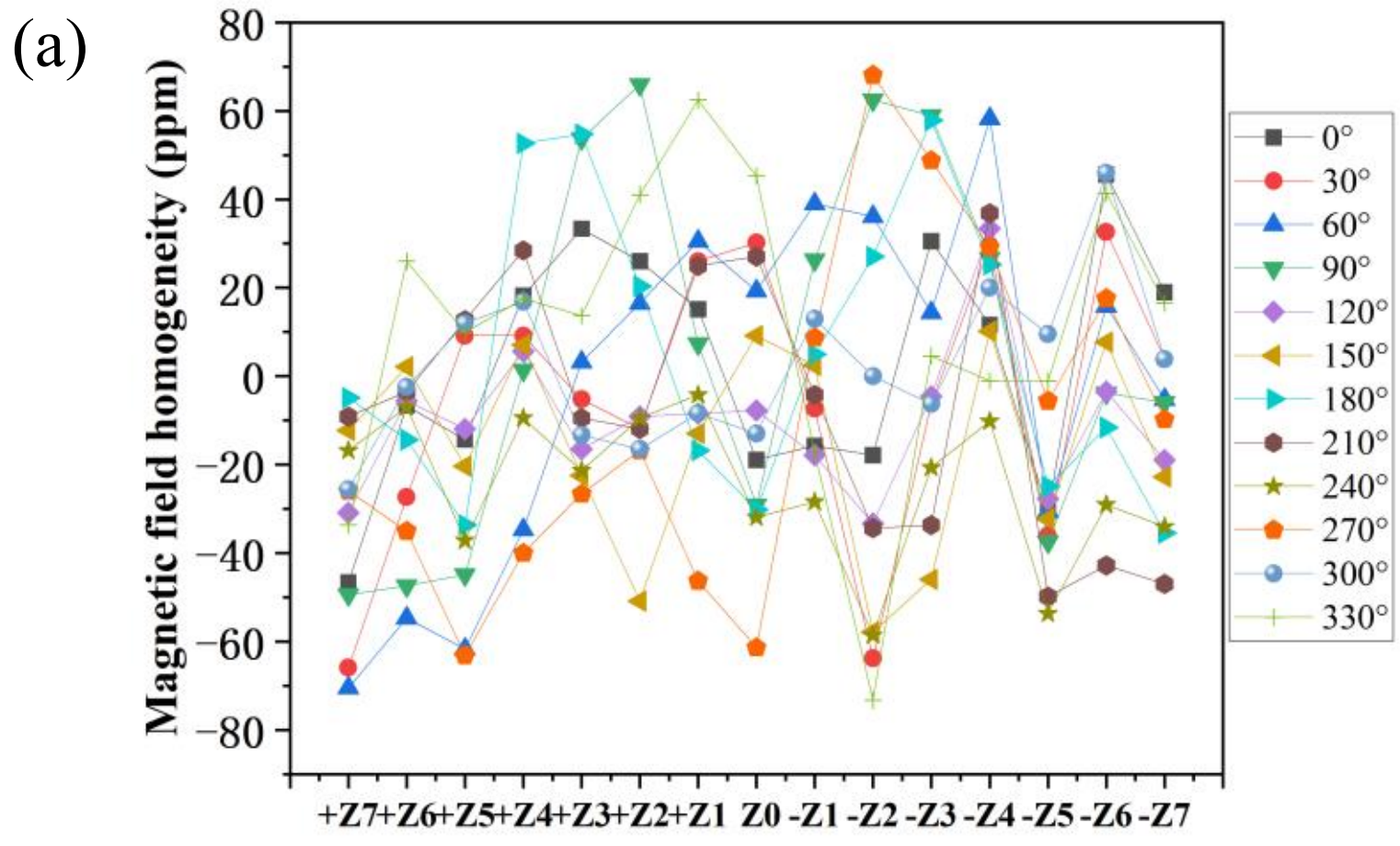

(b)

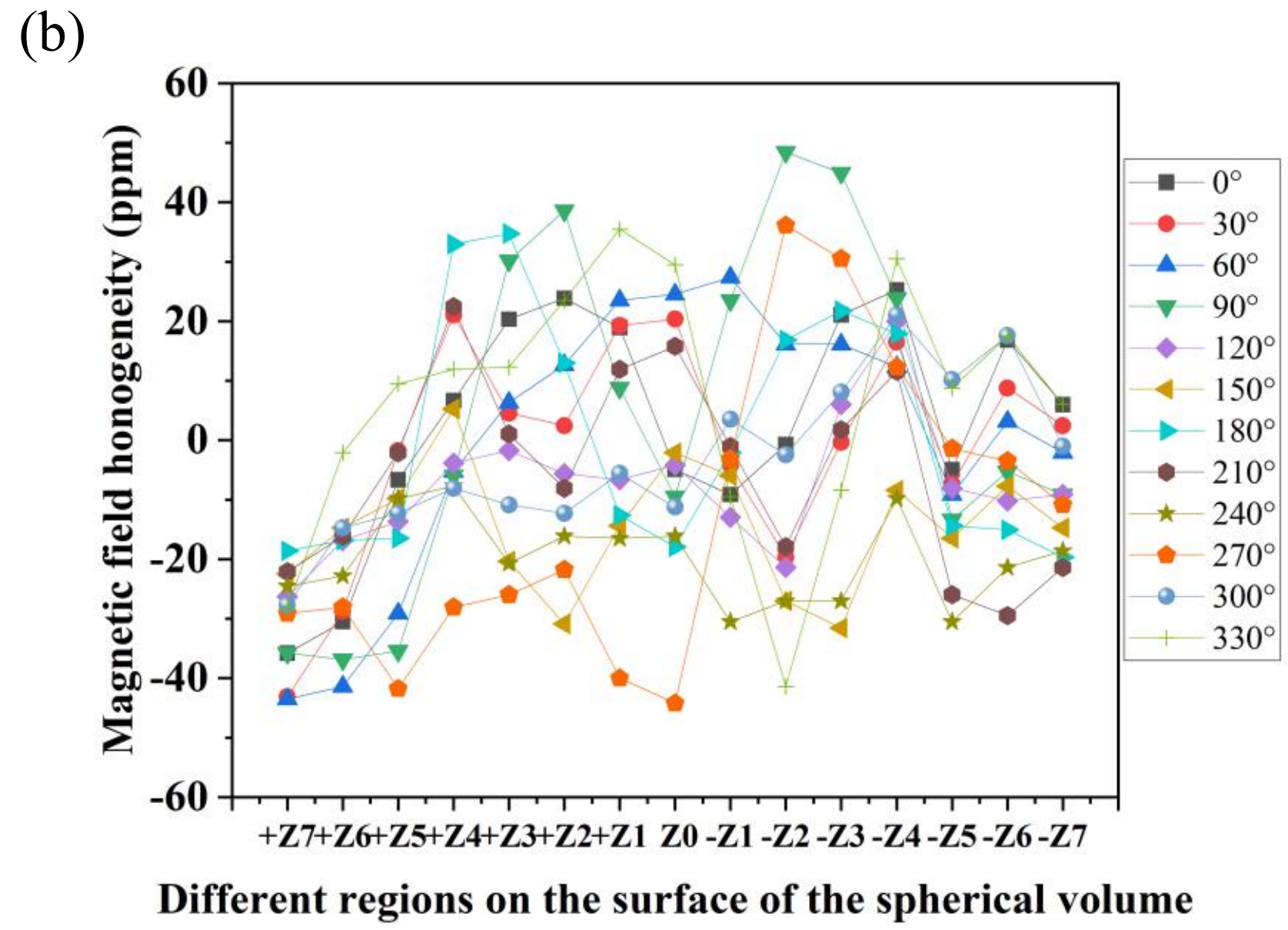

Figure 5. Magnetic-field homogeneity over the surface of (a) 220 mm and (b) 200 mm DSVs sphere as a function of position after passive shimming of the Sm-Fe-N magnet system.

The magnetic field stability of the Sm-Fe-N magnet system was assessed by characterizing the eddy currents induced when the gradient coils were active. The eddy currents in the X and Y directions were very small, at 0.93 ‰ and 1.72 ‰, respectively, while the eddy current along Z was 2.38 %， which is lower than the eddy currents (See Table S1) in low-field Nd-Fe-B and Sm-Co hybrid magnet system without pole pieces. Such small eddy currents originate from the high resistivity of the Sm-Fe-N bonded magnets and from the anti-eddy-current plates. The larger eddy current along Z stems from the low-resistance A3-steel yoke and supports frame with a diameter of 65 cm, which lie along this direction in addition to the Sm-Fe-N magnets. Further increasing the resistivity of the yoke and supports, and applying double-layer self-shielding to the gradient coil in the Z direction would reduce the Z-direction eddy current still more; this work is in progress.

Central magnetic field drift with temperature and time were also measured. It was found that within 4 hours, under a temperature fluctuation ranging from 23.2 ℃ to 23.7 ℃, the maximum amplitude of the central magnetic field variation is $8.9\times10^{-7}$ T, and the relative variation rate of the central magnetic field is 13 ppm.

These results confirm that, because of its high resistivity, the Sm-Fe-N bonded magnet effectively suppresses the eddy currents induced by rapid switching of gradient field. The reduction in eddy currents enables the closely tiled, densely packed magnetic-circuit layout adopted here, which in turn improves both the homogeneity and the stability of the MRI magnet system.

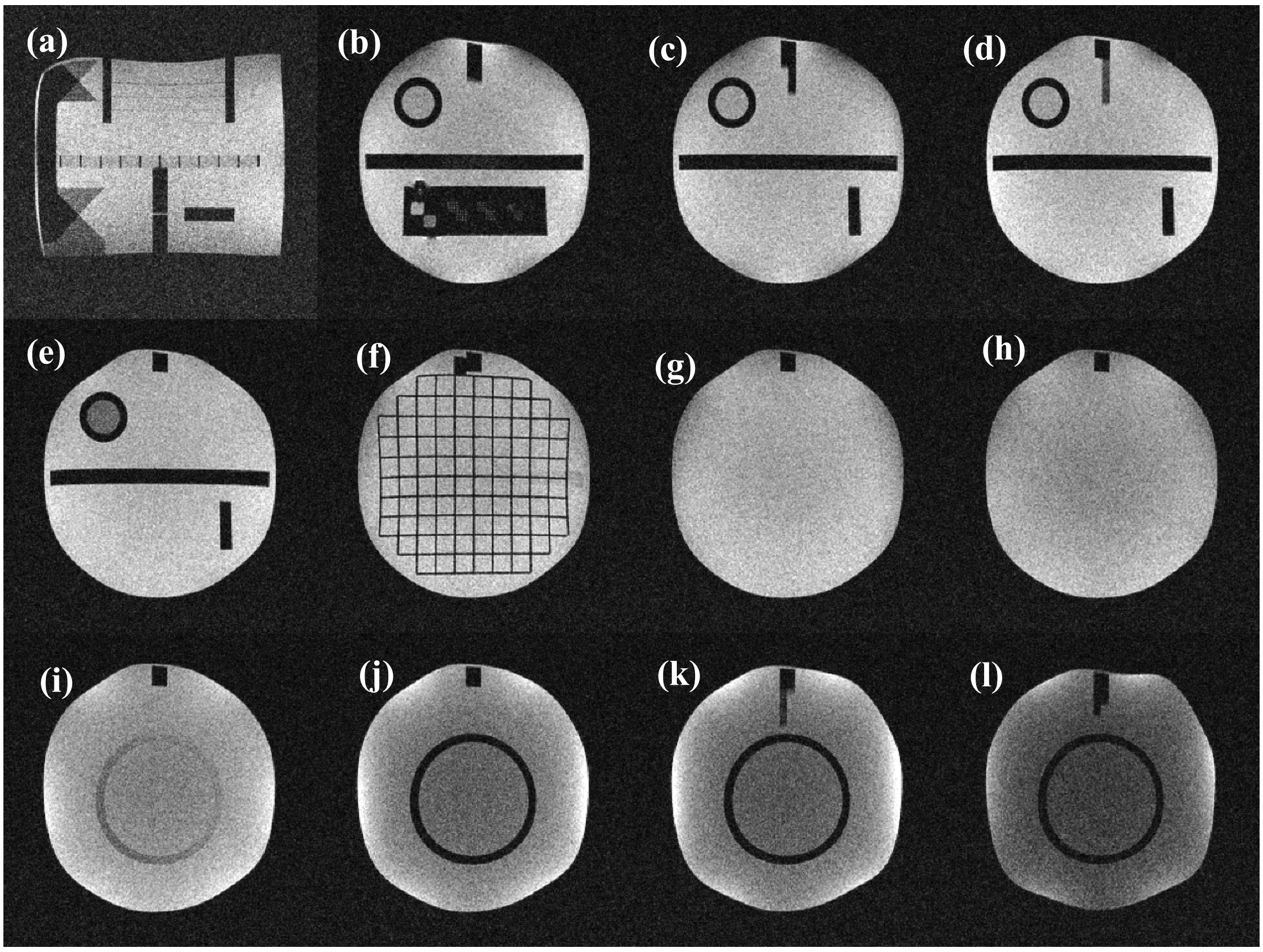

Figure. 6 Quantitative phantom validation on the Sm-Fe-N magnet system using ACR MRI phantom. (a) is the localizer image of the sagittal scan of the ACR MRI phantom using the spin echo (SE) sequence, while (b)-(l) are T1-weighted SE images

from the axial scan. Due to the low SNR of images acquired with a single accumulation on a low-field MRI, we used images with four accumulations to measure parameters such as SNR, CNR, spatial resolution, geometric distortion, and image uniformity.

SNR, CNR, spatial resolution, geometric distortion, and image uniformity of the Sm-Fe-N magnet system were measured using the ACR MRI phantom, and the specific results were shown in Figure 6. Through careful analysis and calculation of the axial T1-weighted SE images, we can conclude that the SNR, spatial resolution and image uniformity of the magnet system are 5.88, 1 mm, and 84.39%, respectively. Regarding geometric distortion, the ACR MRI phantom appears compressed in the horizontal direction, with geometric distortion less than -3.4%; in the vertical direction, it appears stretched, with geometric distortion less than 4.1%. It should be noted that Fig.6 (b) and Fig.6 (c) at the beginning of the water phantom scan, as well as Fig.6 (k) and Fig.6 (l) at the end, exhibit slightly greater distortion compared to the Fig.6 (d-j) in the middle. This is mainly because portions of the two ends of the water phantom (190 mm in diameter and 148 mm in height) extend beyond the DSV (200 mm diameter sphere) of the magnet system's central region, where the homogeneity is high (92.62 ppm). The contrast-to-noise ratio (CNR) could not be detected, which is primarily due to the relatively poor contrast of low-field MRI. It should be noted that the current ACR phantom is primarily used to evaluate the performance of 1.5 T and 3 T MRI systems, and using it to assess the performance of a low-field MRI system may be inappropriate.

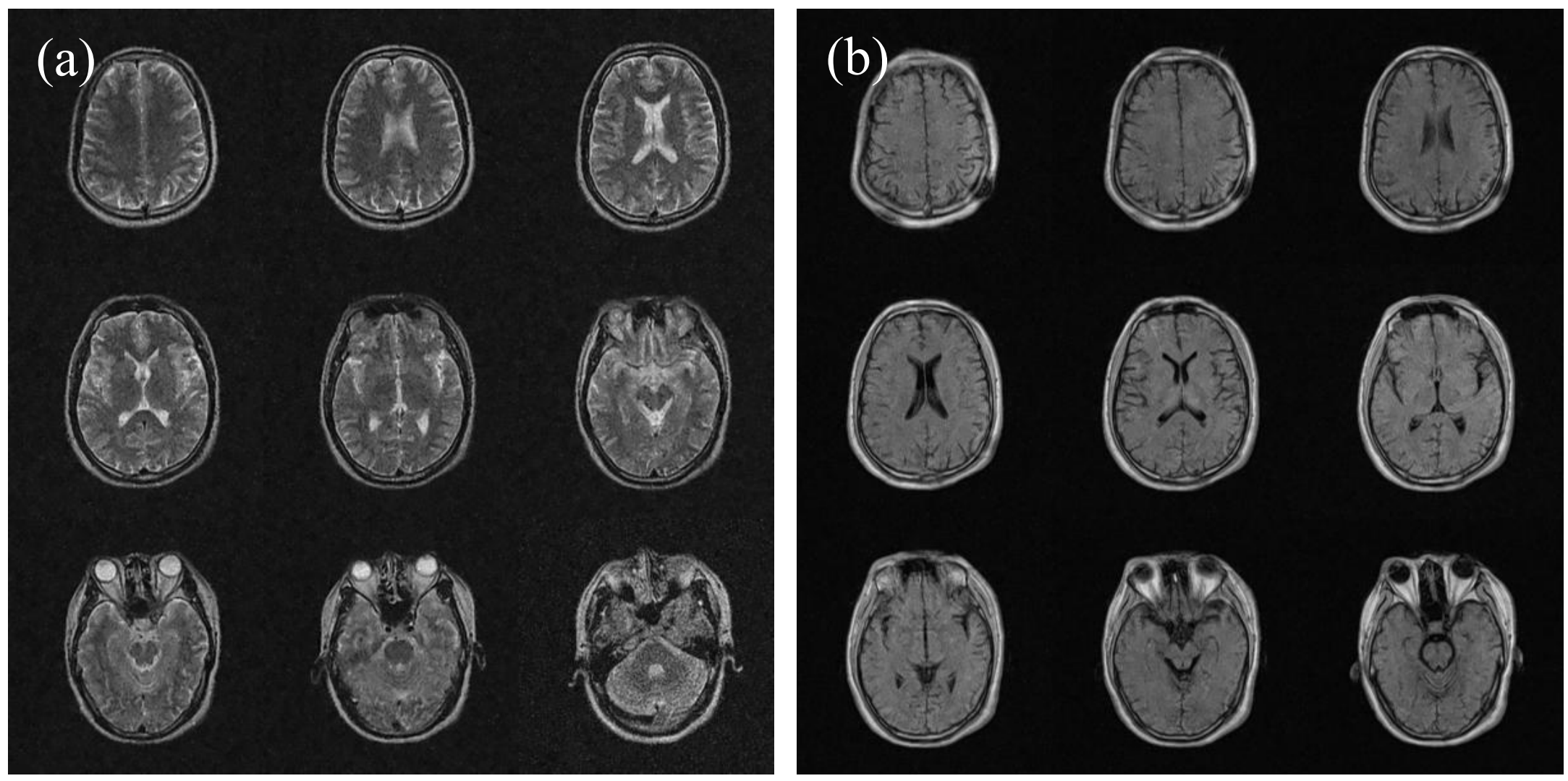

Figure 7. (a) 3D fast-field-echo (FFE) and (b) T2-turbo spin echo (TSE) head images acquired with the Sm-Fe-N MRI magnet system.

Figure 7 shows a 3D fast-field-echo (FFE) head image and a T2-TSE head image of a healthy volunteer. FFE 3D was performed with the following parameters: FOV = 270 mm × 270 mm; in-plane resolution = 1.21 mm × 2.11 mm; slice thickness = 8 mm; number of slices = 16 slices; matrix size = 224 × 128; repetition time (TR) = 44 ms;

echo time (TE) =18 ms; echo train length = 1; total acquisition time = 4 min 36 s. T2-TSE was performed with the following parameters: FOV = 270 mm × 270 mm; in-plane resolution = 1.41 mm × 1.75 mm; slice thickness = 8 mm; number of slices = 15 slices; matrix size = 192 × 154; TR = 5300 ms; TE = 126 ms; echo train length = 11; total acquisition time =5 min 18 s.

The 3D FFE image is essentially free of obvious geometric distortion, indicating good static-field homogeneity in the Sm-Fe-N magnet system; the T2-TSE image shows essentially no banding artifacts, suggesting that eddy currents are indeed small. The absence of obvious distortion or artifacts further confirms that the Sm-Fe-N magnet system delivers low eddy currents and high static-field homogeneity.

## 4. Conclusion

We have developed, for the first time, a low-field, lightweight, Sm-Fe-N MRI magnet system. Compared with sintered Nd-Fe-B and sintered Sm-Co hybrid MRI magnet systems, the Sm-Fe-N bonded-magnet system delivers markedly improved field homogeneity and substantially lower eddy currents: within a 220 mm DSV the field inhomogeneity is below 150 ppm, and the magnitude of eddy current is 0.93 ‰, 1.72 ‰ and 2.38 % along X, Y and Z, respectively. These improvements arise from the high resistivity of the Sm-Fe-N bonded magnet, which enables the closely tiled, densely packed magnetic-circuit layout adopted here. The final imaging results confirm that the Sm-Fe-N magnet system offers high field homogeneity and low eddy currents.

It is noted that although low-field Sm-Fe-N MRI suffers from reduced SNR, spatial resolution, and acquisition speed, it can still enable patient-centric and site-agnostic qualitative or semi-quantitative assessment of gross morphological changes. For example, it can be used for bedside screening and monitoring of intracranial hemorrhage, as well as head imaging screening in low- and middle-income regions and primary healthcare settings.

**Acknowledgment:**

This work was supported by the National Key Research and Development Program of China (No. 2023YFB3507000) and the National Natural Science Foundation of China (No. 52171167).

**Reference:**

1. Fuchs V R and Sox H C Jr 2001 *Health Aff.* **20** 30.
2. Edelman R R and Warach S 1993 *N. Engl. J. Med.* **328** 708.
3. Zhao Y J, Ding Y, Lau V, Man C, Su S, Xiao L F, Leong A T L and Wu E X 2024 *Science* **384** 636.
4. Hennig J 2023 *Magn Reson Mater Phy.* **36** 335.
5. Arnold T C, Freeman C W, Litt B and Stein J M 2023 *J. Magn. Reson. Imaging* **57** 25.
6. Sarracanie M, LaPierre C D, Salameh N, Waddington D E J, Witzel T and Rosen M S 2015 *Sci. Rep.* **5** 15177.
7. Kraus R, Espy M, Magnelind P and Volegov P 2014 *Ultra-low-field Nuclear Magnetic Resonance: A New MRI Regime* (Online: Oxford University Press).
8. Campbell-Washburn A E, Ramasawmy R, Restivo M C, Bhattacharya I, Basar B, Herzka D A, Hansen M S, Rogers T, Bandettini W P, McGuirt D R, Mancini C, Grodzki D, Schneider R, Majeed W, Bhat H, Xue H, Moss J, Malayeri A A, Jones E C, Koretsky A P, Kellman P, Chen M Y, Lederman R J and Balaban R S 2019 *Radiology* **293** 384.
9. Liu Y, Xiao X, Kong X Z, Chen G Q, Zhou J N, Lu F Q, Zhao P F, Pang Y W and Wang Z C 2025 *Med. Phys.* **52** 2874.
10. Kimberly W T, Sorby-Adams A J, Webb A G, Wu E X, Beekman R, Bowry R, Schiff S J, de Havenon A, Shen F X, Sze G, Schaefer P, Iglesias J E, Rosen M S and Sheth K N 2023 *Nat. Rev. Bioeng*. 1 617.
11. Mazurek M H, Cahn B A, Yuen M M, Prabhat A M, Chavva I R, Shah J T, Crawford A L, Welch E B, Rothberg J, Sacolick L, Poole M, Wira C, Matouk C C, Ward A, Timario N, Leasure A, Beekman R, Peng T J, Witsch J, Antonios J P, Falcone G J, Gobeske K T, Petersen N, Schindler J, Sansing L, Gilmore E J, Hwang D Y, Kim J A, Malhotra A, Sze G, Rosen M S, Kimberly W T and Sheth K N 2021 *Nat. Commun.* **12** 5119.
12. Lother S, Schiff S J, Neuberger T, Jakob P M and Fidler F 2016 *Magn. Reson. Mater. Phy.* **29** 691.

13. Zhang Y X, Kong X H, He W and Xu Z 2024 *IEEE Trans. Instrum. Meas.* **74** 4003909.

14. Liu Y L, Leong A T L, Zhao Y J, Xiao L F, Mak H K F, Tsang A C O, Lau G K K, Leung G K K and Wu E X 2021 *Nat. Commun.* **12** 7238.

15. Reilly T O, Teeuwisse W M, Gans D, Koolstra K and Webb A G 2021 *Magn. Reson. Med.* **85** 495.

16. Cooley C Z, McDaniel P C, Stockmann J P, Srinivas S A, Cauley S F, Śliwiak M, Sappo C R, Vaughn C F, Guerin B, Rosen M S, Lev M H and Wald L L 2021 *Nat. Biomed. Eng.* **5** 229.

17. Lee P K, Qiu Y Q, Wang C Y and Zhang Z Y 2026 *Magn. Reson. Med.* 95 188.

18. Wei S F, Wei Z, Wang Z, Wang H X, He Q Y, He H Y, Li L and Yang W H 2023 *Magnetic Resonance Materials in Physics, Biology and Medicine* **36** 409

19. Shen P, Guo J Q, Han J Z, Zhang Z H, Chen X G, Yang W Y, Liu J, Ji X H, Zhou D and Yang J B 2026 *Chinese Phys. B* **35** 057503.

20. Poole M S and Hugon C, “Low-Field Magnetic Resonance Imaging Methods and Apparatus” U.S. Patent No. 16/742.311， issued May 14， 2020.

21. Zhang T L, Zhang B, Wang H, He Y K, Xu C, Wang X Q, Zhang W, Zhang Z H and Jiang C B 2018 *J. Magn. Magn. Mater.* **466** 38.

22. Zhang X F, Liu L B, Li Y Q, Zhang D T, Liu W Q and Yue M 2024 *Chinese Phys. B* **33** 097503

23. Shen P, Qian H D, Han J Z, Liu T, Yan Z, Ji M, Zhou L, Liu W Q, Liu S Q, Yang W Y, Li Y, Cao H H, Yue M, Yang J B and Yang Y C 2025 *Chinese Phys. B* **34** 087503

# Supplementary Materials: magnet specifications, assembly procedure and comparison with Nd-Fe-B+ Sm-Co hybrid magnet system

## 1. Magnet specifications

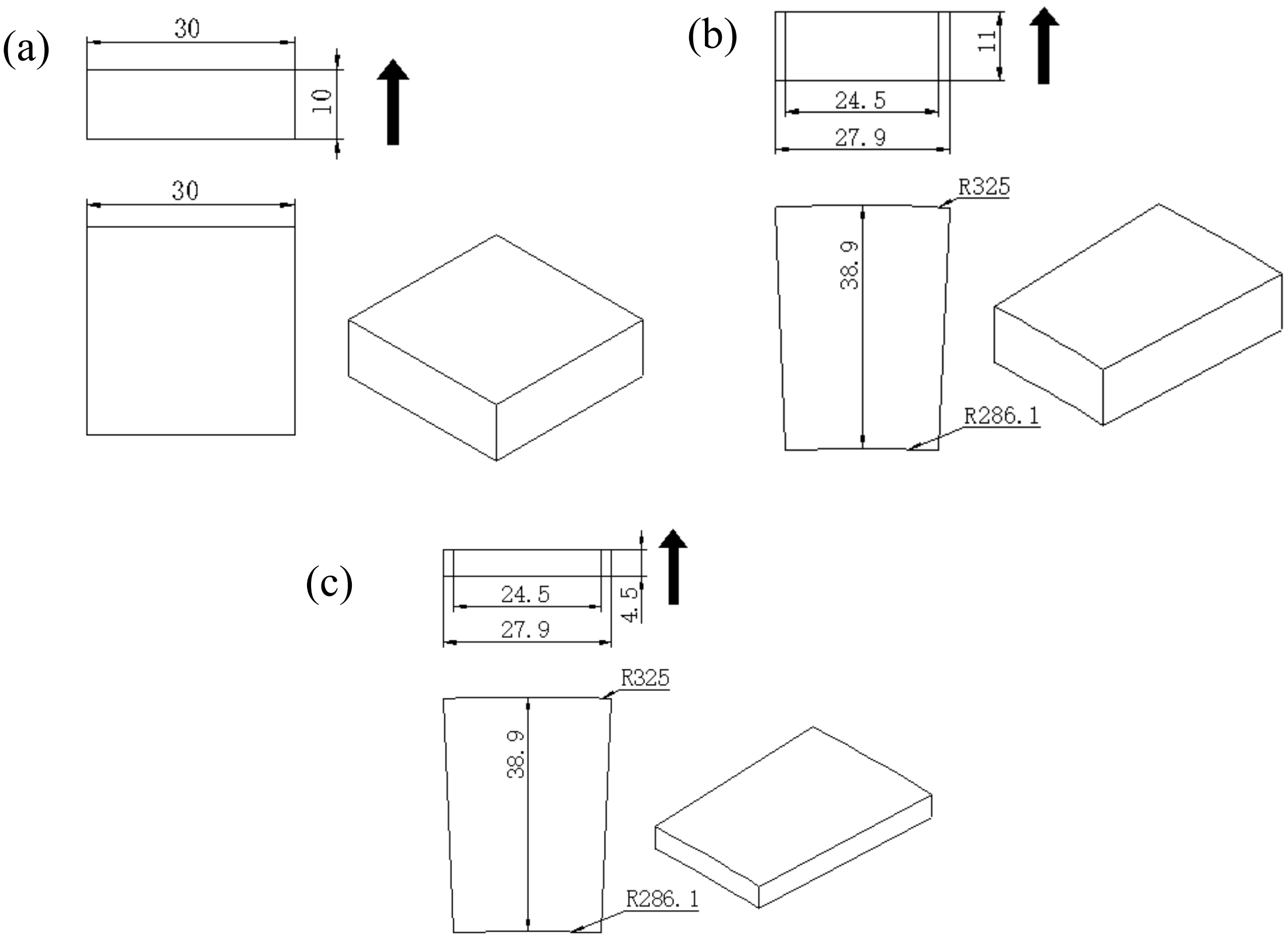

Figure S1. Specifications and dimensions of the three injection-molded magnets.

## 2. Assembly procedure

Assembly procedure of the Sm-Fe-N magnet system as follows: Square magnets of 30 mm × 30 mm × 10 mm (Figure S1(a)) were stacked along the thickness direction (10 mm) into three layers, giving a total thickness of 30 mm, and then placed inside the pre-assembled iron-yoke support (Figure 2). The periphery of the tiled assembly was machined to form a circular disk with an outer diameter of 65 cm and a thickness of 30 mm. Sector-shaped magnets 11 mm thick (Figure S1(b)) were stacked into three layers along the thickness direction to reach 33 mm and assembled into a ring with an outer diameter of 65 cm and a thickness of 33 mm. This 33 mm ring was placed on the 30 mm circular disk, aligned by their outer diameters, so that the outermost rim of the assembled magnet block reached 63 mm in height. On the left and right edges of the outermost rim of each pole disk, an additional 4.5 mm sector-shaped magnet (Figure S1(c)) was added on top of the 63 mm stack, bringing the left- and right-side thickness of each pole disk to 67.5 mm while keeping the front and back sides at 63 mm. The upper and lower pole disks share an identical magnet-stacking pattern and are kept in mirror symmetry during installation. The two assembled pole disks form the N and S poles, respectively.

### 3. Comparison with Nd-Fe-B+ Sm-Co hybrid magnet system

Table S1 Comparison of specifications, dimensions, and performance parameters between Nd-Fe-B+ Sm-Co hybrid magnet systems and Sm-Fe-N magnet systems.

| magnet system | magnetic field strength | Magnet layout | shimming strategy | DSV | Homo-geneity | gradient design | eddy currents (echo time difference [2]) | Geo-metry |
|---|---|---|---|---|---|---|---|---|
| Hybrid: Nd-Fe-B and Sm-Co | 0.07 T | dispersed ring-shaped layout [1] | Passive shimming | 200 mm | 677.47 ppm | Gx and Gy gradient coils without shielding, Gz gradient coil with single-layer shielding | X: 360 μs<br>Y: 640 μs<br>Z: 4560 μs | [3]PDD: 580 mm<br>[4]PDG: 300mm |
| Sm-Fe-N | 0.067 T | densely packed layout | Passive shimming | 200 mm | 92.62 ppm | Gx and Gy gradient coils without shielding, Gz gradient coil with single-layer shielding | X: 92 μs<br>Y: 224 μs<br>Z: 3536 μs | [3]PDD: 650 mm<br>[4]PDG: 300mm |

dispersed ring-shaped layout [1]: a 5 mm gap is maintained between magnets of different rings, while a 3 mm gap is kept between magnets within the same ring for the hybrid magnet system for decreasing the eddy current.

echo time difference [2]: The Sm-Co and Nd-Fe-B hybrid magnet system uses an MR solution spectrometer, while the Sm-Fe-N magnet system uses a Firstech spectrometer. Although the Sm-Fe-N magnet system can obtain the eddy current gradient field generated after the application of a gradient pulse magnetic field, and thereby calculate the percentage of the eddy current field relative to the gradient pulse field, the hybrid magnet system cannot obtain the eddy current gradient field due to differences in software and hardware. To facilitate the comparison of eddy currents between the two magnet systems, we adopted a method of measuring the time difference between the time when the spin echo generated by applying a positive gradient pulse magnetic field reaches its maximum and the time when the spin echo generated by applying a negative gradient pulse magnetic field reaches its maximum, in order to assess the magnitude of the eddy currents. We measured the time difference between the time when the spin echo generated by applying a positive gradient pulse reaches its maximum and the time when the spin echo generated by applying a negative gradient pulse magnetic field reaches its maximum along the X, Y, and Z directions, respectively. Each direction was measured three times, and the average values were taken and filled into the Table.4. The smaller the time difference, the smaller the eddy currents; conversely, the larger the time difference, the larger the eddy currents. As can be seen from Table 4, the time differences of the echoes generated by the Sm-Fe-N magnet system are smaller than those generated by the hybrid magnet system in all three directions, X, Y, and Z. Therefore, it can be concluded that the eddy currents of the Sm-Fe-N magnet system are smaller than those of the hybrid magnet system in all three directions, X, Y, and Z.

[3]PDD: pole disk diameter; [4]PDG: pole disk gap;